# Who Let the Diamonds Out?

Vincent Halde[*], Olivier Bernard[*], Mathieu Brochu[*], Laurier Dufresne[*], Nicolas Fleury[*], Kayla Johnson[*], Benjamin Moffett[*], David Roy-Guay[*]

[*]*SBQuantum, 805 Rue Galt O, Sherbrooke, Quebec J1H 1Z1, Canada*

Nitrogen-Vacancy (NV) center magnetometry is a highly promising quantum sensing technology, with early prototypes demonstrating impressive sensitivity in compact sensing heads. Yet, most existing implementations remain tied to laboratory setups, lacking the portability and environmental robustness needed to unlock their full potential in real-world applications. In this work, we introduce a fully portable, hand-held NV-based magnetometer that delivers a vector sensitivity of approximately 400 pT/√Hz, heading errors below 5 nT in Earth's field, and a wide signal bandwidth that supports on-field recalibration and operation on moving platforms. We further demonstrate the system's technological maturity through environmental qualification such as thermal, vibration, radiation and other operational stresses related to deployment in low Earth orbit, and through successful deployments in demanding scenarios, including northern Canadian weather conditions, drone-mounted surveys and high-altitude balloon flights. Together, these achievements establish this NV-based magnetometer as a robust, versatile tool ready to bring quantum sensing performance to a broad range of field and autonomous applications.

## 1. Introduction

Nitrogen-Vacancy (NV) center magnetometry has the potential to enable magnetic field measurements with high sensitivity, broad dynamic range, and operation in extreme conditions (pressure, temperature). NV centers, atomic-scale defects in diamond, can be optically initialized and read out, allowing vector magnetic fields to be measured through spin-resonance techniques. Because of these properties, NV-based magnetometers are attracting growing interest across various applications, from fundamental physics and biomedical imaging to geophysical exploration and space-based Earth observation.

Despite their potential, most current NV implementations are not yet practical for real-world deployment. High-performance demonstrations often feature miniature sensing heads connected to bulky laboratory equipment, making them unsuitable for mobile or volume-restricted platforms such as drones, small satellites, or handheld systems. Furthermore, many experimental systems are limited to single-axis measurements[1–6], failing to leverage the full vector measurement capability inherent to the NV defect's crystalline symmetry.

To unlock the true potential of NV magnetometry, it is essential to move beyond laboratory prototypes toward fully integrated, portable, vector systems. Our work at SBQuantum aims to do precisely this by exploiting the NV center's combination of high sensitivity and intrinsic absolute vector accuracy, a type of measurement traditionally achievable only by pairing multiple sensor types, with significant penalties in size, weight, and power (SWaP)[7,8]. Alternative approaches, such as vectorizing scalar OPMs, impose unacceptable sensitivity trade-offs and are therefore not considered commercially viable solutions for the moment[9] .

This paper presents SBQuantum's progress toward achieving the trifecta of sensitive, accurate, and truly vector magnetometry in a compact, field-ready device. We present the design of the diamond-based magnetometer, provide an overview of its magnetic performance, and demonstrate its practical usability through real-world field deployment scenarios.

## 2. SBQuantum Diamond Magnetometer

The SBQuantum Diamond Magnetometer (SBQDM) is built on NV centers in single-crystal diamond. These atomic-scale defects exist in four distinct crystallographic orientations and are mostly sensitive to the magnetic field component aligned with its NV axis. By probing a bulk diamond containing trillions of NV centers distributed among these orientations, the SBQDM measures the field along four projections, from which it reconstructs the full three-component magnetic vector, as well as the diamond's temperature[10,11].

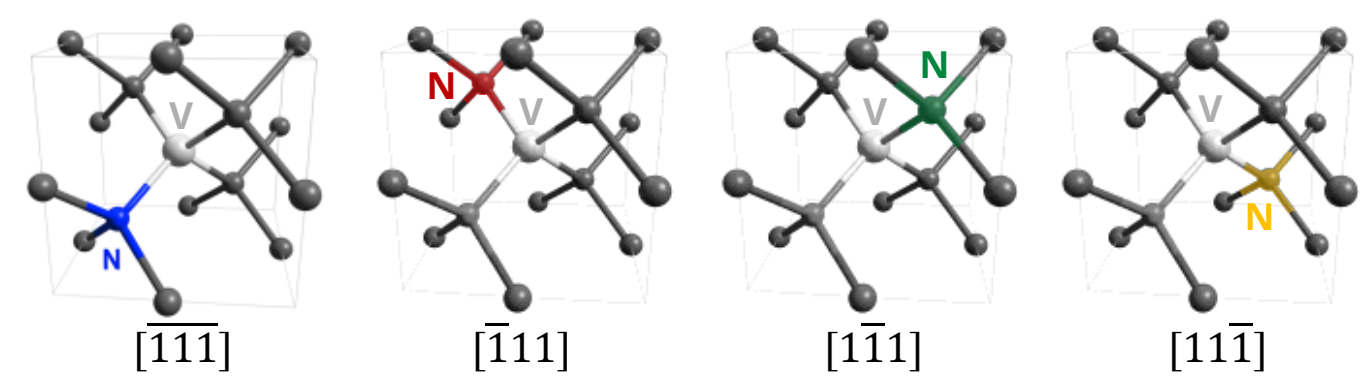

*Figure 1: Four different configurations of the NV center defect in the crystal lattice. Each configuration is mostly sensitive to the magnetic field along its NV axis.*

The magnetic measurement process is based on Optically Detected Magnetic Resonance (ODMR) and consists of 3 steps: <u>initialization</u>, <u>manipulation</u>, and <u>readout</u>.

The <u>initialization</u> of NV centers into their ground state is made through optical pumping with green laser light. The <u>manipulation</u> of the spin state is performed via microwave frequencies resonant with the spin transitions to drive populations between states. The <u>readout</u> of the system is carried out by measuring the red photoluminescence of the diamond under green light excitation, since NV centers exhibit reduced photoluminescence when in the magnetically sensitive $m_s = \pm 1$ states compared to the non-sensitive ground state [11].

To independently address each NV orientation and extract vector information, a bias magnetic field is applied at the diamond. This field, produced by small permanent magnets, ensures each axis has a distinct resonance frequency, allowing selective microwave addressing. Any external magnetic field then causes measurable frequency shifts on each orientation, enabling precise vector reconstruction.

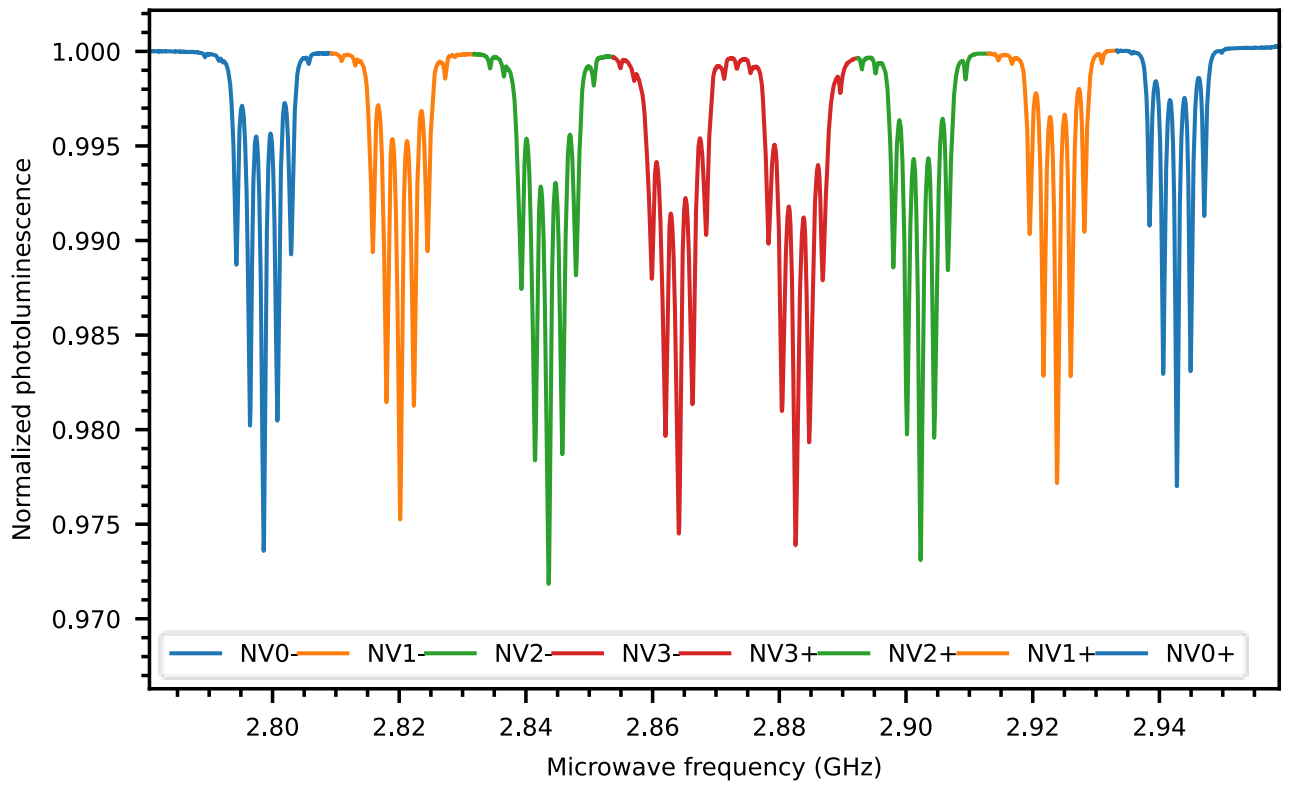

*Figure 2: ODMR spectrum taken on a SBQDM. The 8 peaks represent the 4 sensing axis pairs, colored for visual identification. Magnetic field along each sensing axis will modify the splitting between each element of a pair, allowing for vector reconstruction.*

The SBQDM integrates all ODMR subsystems into a self-contained instrument with no need for external laboratory equipment, using commercial off-the-shelf (COTS) electronic and optics components. An example of ODMR produced by the device is shown in Figure 2. Figure 3 shows that every printed circuit board in the SBQDM corresponds to a typical NV center magnetometry laboratory equipment, stripped to its minimum requirements. This approach delivers a miniature form factor with reduced power consumption while maintaining high performance.

The elongated form factor of the SBQDM reflects a design philosophy centered on maximizing magnetic accuracy. To meet stringent (nanotesla scale) performance requirements, the instrument minimizes magnetic disturbances at the sensing region by strategically distancing potential sources of interference. Most COTS electronic components contain ferromagnetic or paramagnetic materials and are placed as far as possible from the sensor head. Additionally, electrical currents are routed with precision to limit their magnetic influence, ensuring that both static and dynamic perturbations are kept well below the noise floor of the sensing element.

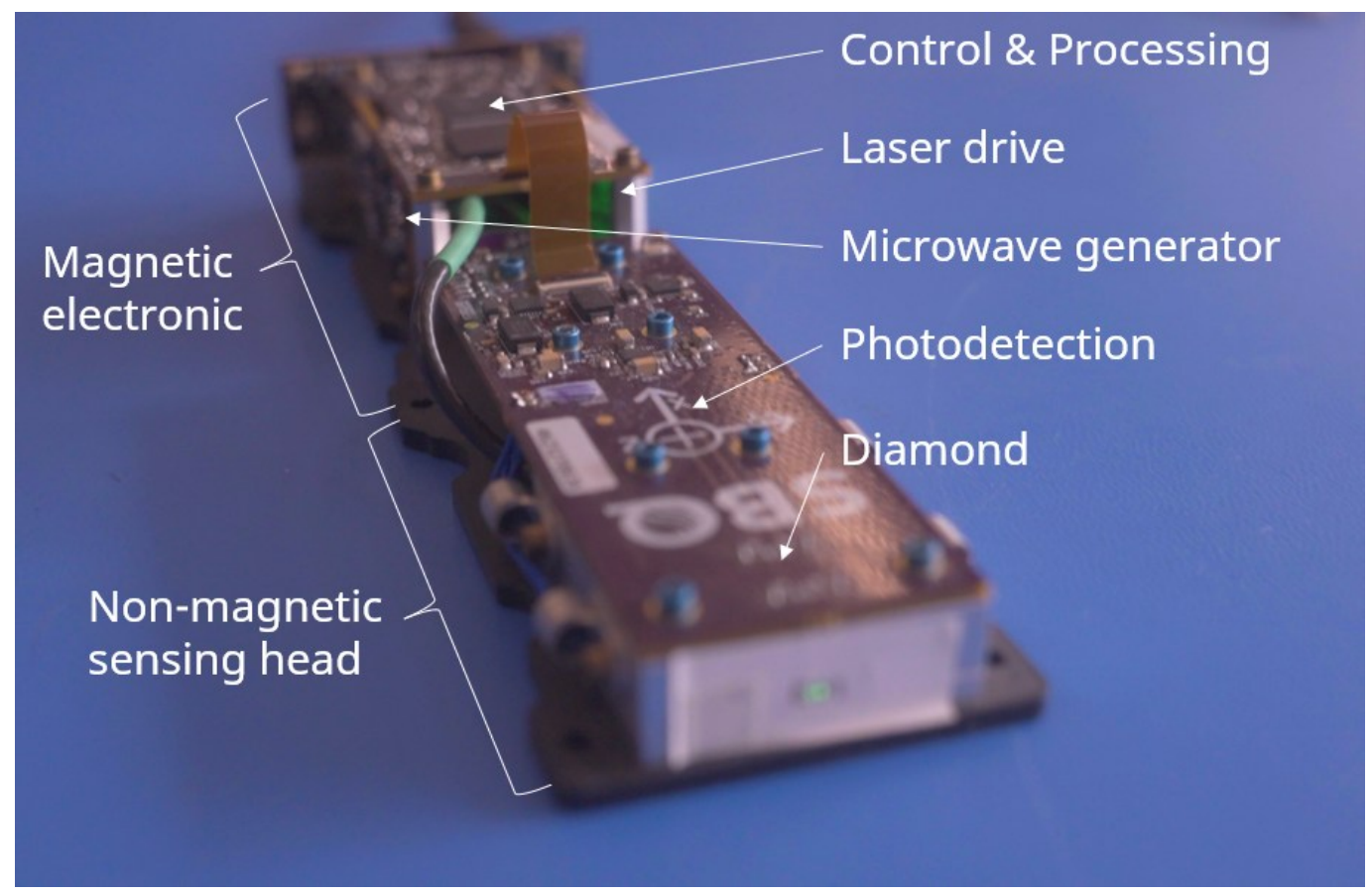

*Figure 3: SBQDM without its external casing, showing the submodules that make the experiment possible*

Although the bias magnets are manufactured from high-stability materials, residual thermal effects can produce small offset drifts. These effects are mitigated by anchoring the magnets thermally to the aluminum frame of the sensing head and actively monitoring their temperature. Each unit undergoes an in-house calibration procedure to correct for residual temperature dependencies and assembly variations.

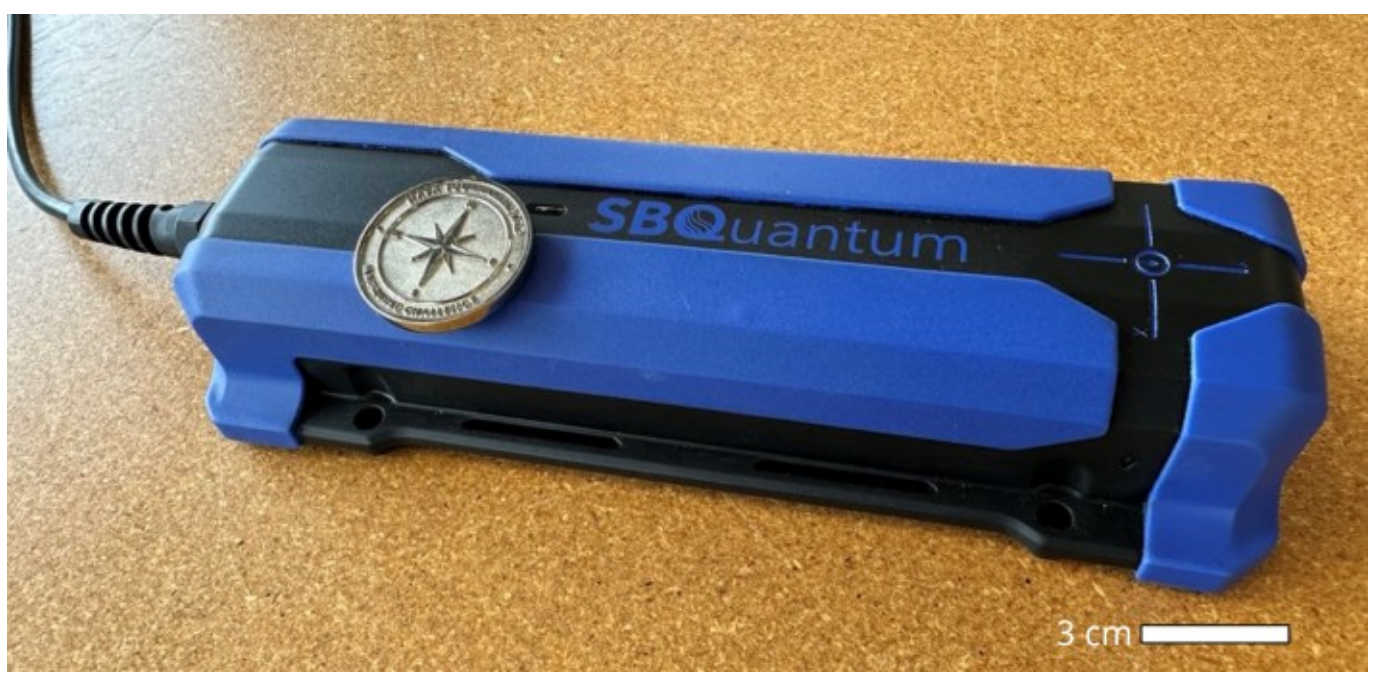

*Figure 4: The SBQDM has USB and power connection. Most of the processing is done internally, allowing for real-time magnetic field measurement.*

The resulting instrument is compact and portable, measuring only 23 × 7 × 4 cm (including casing), weighing 483 g, operating at under 6 W at 24 V, and USB connected. It provides real-time vector magnetic measurements without any external processing or equipment, while machine-learning correction algorithms are applied locally on the host computer (see Figure 5). This combination of quantum sensing principles and engineering optimization makes the SBQDM a portable, robust solution for high-precision magnetometry in real-world environments.

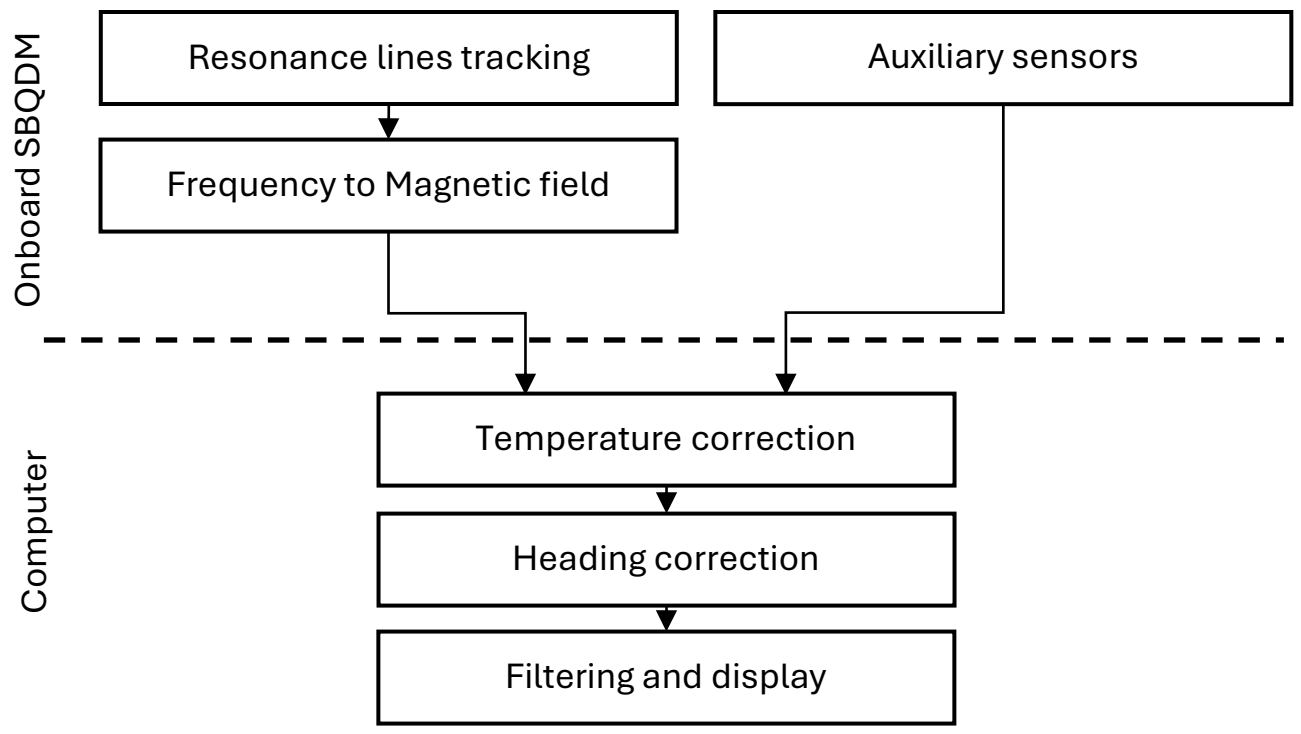

*Figure 5: Processing pipeline of the SBQDM. The eight resonance lines are tracked onboard the SBQDM and the peak frequencies are converted to vector magnetic field measurements. The SBQDM also acquires auxiliary sensors (temperature, currents, voltages) that feed in the ML correction algorithm. Two layers of corrections can be applied in real time on the computer: temperature correction and heading error correction. The data is then displayed, with optional filtering.*

## 3. Performances validation

### *3.1 Sensitivity*

The measure of sensitivity is performed with the device placed inside a zero Gauss chamber (Magnetic Shield Corp. ZG-206), which shields the sensor from external magnetic field perturbations. This shielding is critical, as SBQuantum's headquarters are located near a steel-beam manufacturing facility, where passing trucks and nearby vehicles can easily be detected in unshielded measurements even from the third floor, an example of which is shown in Figure 6.

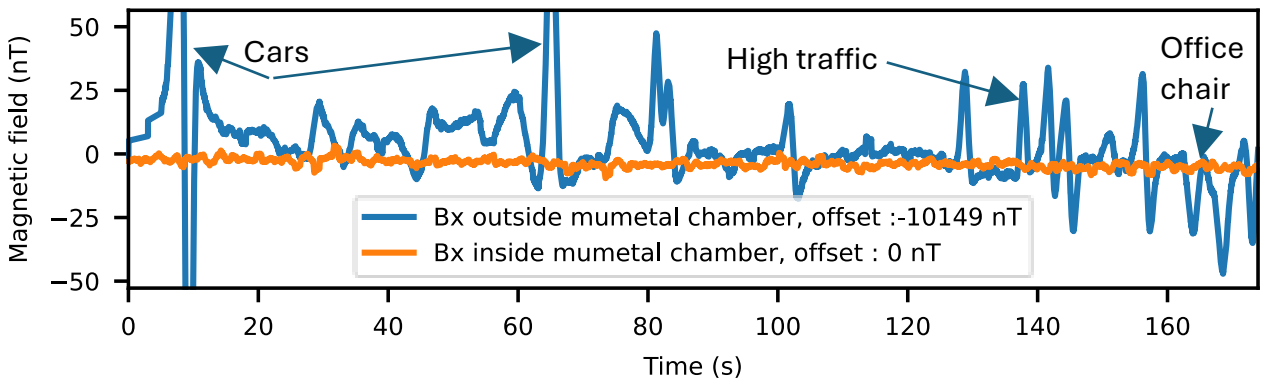

*Figure 6: Demonstration of shielding of the zero Gauss chamber. Blue: magnetic acquisition outside of the zero Gauss chamber, showing magnetic signal from cars and office chairs. Orange: magnetic signal at the same location but inside the zero Gauss chamber.*

For the test, the device was allowed to thermally stabilize for 20 minutes before recording a 2 minutes magnetic field acquisition. The resulting data were analyzed using Allan deviation, revealing a constant noise floor up to approximately 5 seconds of averaging time, beyond which thermal drift dominates as the readings were not corrected for temperature effects for this device. Measured sensitivities at 1 Hz are [346, 434, 401] $pT/\sqrt{Hz}$ for the Bx, By, Bx axis respectively. Theoretically, according to the shot-noise limit of the system, the device could achieve sensitivity levels down to 12 $pT/\sqrt{Hz}$. The difference between this theoretical limit and the measured values is attributed to residual noise sources, including laser intensity fluctuations, microwave instability, and other minor sources of magnetic interference within the instrument.

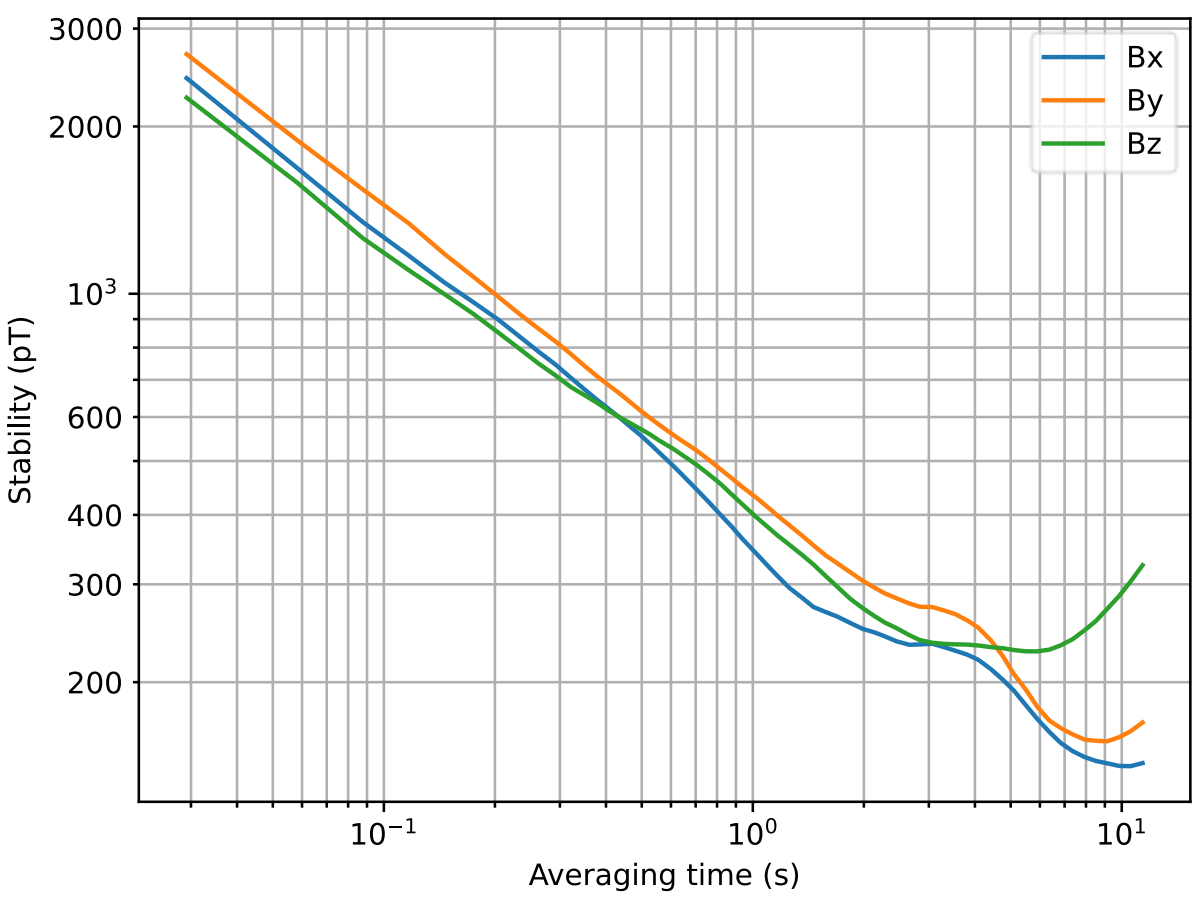

*Figure 7: Allan deviation of a magnetic acquisition in a zero Gauss chamber. The graph shows a steady improvement on stability with averaging time, indicator of white noise regime down to ~5 seconds, where flicker noise and drift contribution take over. Some of this drift contributions can be removed through a temperature trained ML model (not applied here).*

### *3.2 Accuracy*

One of the less frequently highlighted advantages of NV-center-based magnetometry is its inherent potential for absolute vector accuracy. Indeed, the orthogonality of the sensing axes is defined and fixed by the diamond's crystalline lattice, and the scale factor and linearity are determined solely by the Hamiltonian inversion once the resonance line frequencies are identified. One major challenge remains to control the offset created by the permanent magnets or coils of the typical ODMR scheme. The SBQDM mitigates these effects by integrating multiple temperature sensors and other auxiliary sensors to correlate offset variations with environmental changes, enabling effective post-processing compensation.

Accuracy characterization was conducted at Natural Resources Canada's (NRCan) Ottawa Geomagnetic Observatory, where reference coils provide fields with an absolute accuracy under 1 nT. A training dataset of 379 points was first recorded to allow an internal machine-learning model to correct non-linearities in the inversion process and compensate for perturbations caused by magnetic components in the device. A test dataset of 253 randomly distributed points, spanning amplitudes from 0 to 60 µT, was then used to evaluate the calibrated performance.

The results show a standard deviation of the error under 5 nT across the test set. Calibration parameters such as the scale factors, orthogonality and offset can be extracted from this analysis. Table 1 confirms that ML calibration can correct for both scale and orthogonality close to ideal values, while offset stability remains limited to nT level.

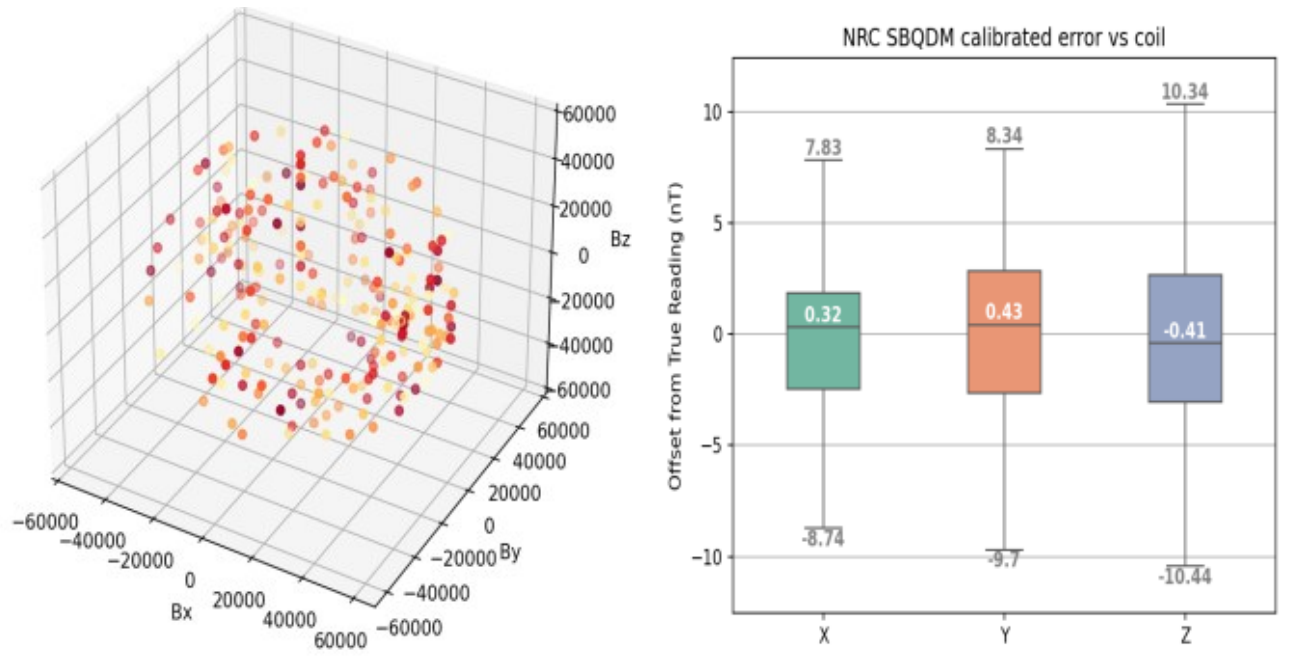

*Figure 8 (left) Accuracy testing set with 253 points taken at random values within the operating range of the sensor. (right) Residual when compared to laboratory reference, for each axis.*

*Table 1: Calibration parameters before and after ML compensation*

| | | Scale | Orthogonality (rad) | Offset (nT) |
|---|---|---|---|---|
| Before calibration | X | 1.00323 | 860E-5 | 314 |
| | Y | 1.00950 | 336E-5 | -86 |
| | Z | 1.00209 | 115E-5 | -2244 |
| After calibration | X | 0.99998 | 4E-5 | 2.1 |
| | Y | 0.99997 | -9E-5 | 4.2 |
| | Z | 1.00014 | -20E-5 | 2.1 |

The primary source of offset instability arises from bias field and other perturbators' behaviors not fully captured by the machine-learning training set. Additional contributors include magnetization of nearby components during handling or transport and slow variations in resonance line shape caused by laser and microwave power fluctuations. Nevertheless, offsets can be further reduced through frequent on-field recalibration (see Figure 9), and other techniques are being developed and tested by SBQuantum, one of which involves removing the bias field altogether [12].

Non-linearities were also evaluated by applying controlled ramps from a calibrated coil. On a different unit than the one reported in Figure 8, a ramp from –50 μT to +50 μT yielded a standard deviation of 7.5 nT.

### 3.3 Bandwidth

The bandwidth of a magnetic sensor is a critical specification when measuring dynamic signals or operating across a large dynamic range, such as Earth's magnetic field. To maintain nanotesla-level accuracy while rotating through Earth's field, where a single vector component can vary by up to 100 μT, the sensor must have a bandwidth cutoff well above the highest frequency of interest. Without sufficient bandwidth, sensitivity and accuracy metrics become irrelevant once the sensor is deployed in the field on mobile platforms.

Figure 9 demonstrates the SBQDM's ability to sustain a hand-executed spin calibration routine with rotation periods of about 10 seconds. To evaluate dynamic performance, the Total Magnetic Intensity (TMI) was monitored, which should remain constant in this controlled test environment. In the raw data (blue trace), TMI variations on the order of 1 μT are observed, primarily due to residual offsets not calibrated in this new environment. Applying the spin calibration[13] mainly removes constant offsets on each vector component, reducing these variations significantly. After correction, the TMI fluctuations are limited to within 20 nT, corresponding to a dynamic heading error below 20 nT.

This capability confirms that the SBQDM maintains high accuracy during dynamic motion when properly calibrated. The spin calibration feature is therefore an essential tool for field deployment, allowing the offset to be verified and corrected both before and after surveys, ensuring reliable performance even in challenging, dynamic conditions.

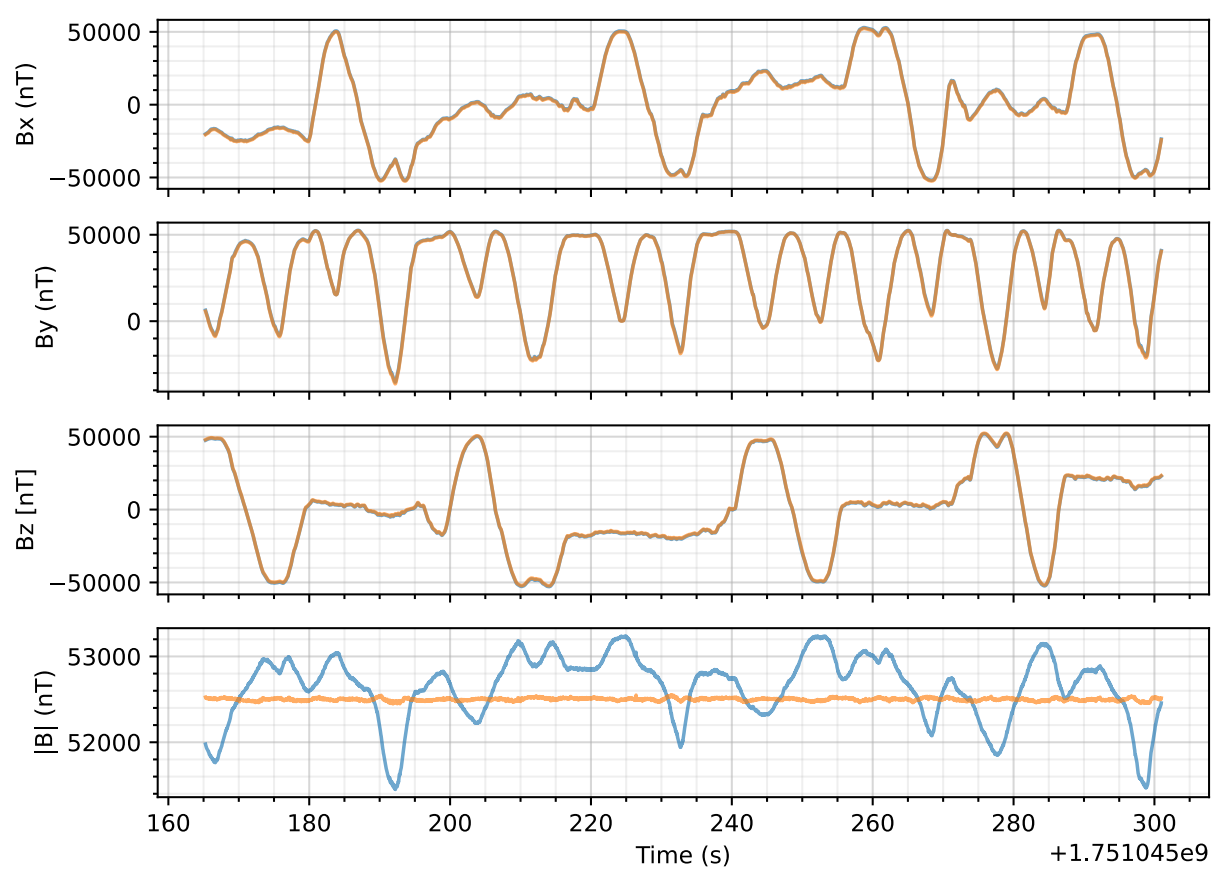

*Figure 9: Hand-made spin-calibration in a constant magnetic field environment constrains the measurement to a flat TMI, resulting in offset, scale and orthogonality correction that can be executed on the field prior to a measurement. Blue: raw magnetic field data. Orange: corrected magnetic field data with spin-calibration assuming a constant TMI*

## 4. Field Deployment

We demonstrate the field readiness of the SBQDM through deployments in challenging environments and use cases where the advantages of NV-center magnetometry offer clear benefits, such as vector accuracy, stability, and robustness.

*4.1 Space deployment*

The development of the SBQDM was originally driven by the MagQuest Challenge[14] organized by the U.S. National Geospatial-Intelligence Agency (NGA). This challenge sought new magnetic sensing technologies to contribute to the World Magnetic Model (WMM), which is currently derived from data collected by ESA's SWARM satellite constellation[15].

To ensure that the SBQDM could survive and operate in space, the instrument was subjected to rigorous environmental testing replicating satellite launch and operational conditions. The survivability ranges validated through these tests are summarized in Table 2.

*Table 2: Environmental condition test for the SBQDM*

| Parameter | Test name | Survivability range |
| --- | --- | --- |
| Temperature | High Temperature | 125 °C for 335 hours |
| | Low Temperature | -65 °C for 335 hours |
| | Temperature Cycling | 335 cycles -40 to 80 °C |
| Pressure | TVAC | 5 x 10-5 Torr at 50 °C for 24 hours |
| Mechanical | Sinusoidal | 5 to 13 Hz at 0.5 $in_{pp}$<br>13 to 2000Hz at $4.5g_{peak}$ |
| | Acceleration | 12g for 20s |
| | Impact | 50g $g_{peak}$, 0.1 ms |
| Radiation | Proton | 40 krad at 105 MeV |
| | Gamma | 46 krad |

These tests primarily focused on mechanical and thermal endurance rather than full operational characterization, as the strong magnetic signatures of test equipment preclude accurate performance validation under these extreme conditions. While some performance degradation was observed in certain test conditions, these insights have been used to refine the design, further improving long-term robustness and space readiness.

*4.2 Extreme conditions deployment*

Magnetometers are essential in mineral exploration, where they provide a cost-effective, non-invasive means of mapping subsurface features before drilling. The Canadian North, with its high mineral potential, provides an ideal proving ground for instruments designed to withstand extreme environments.

To validate the SBQDM's durability and stability, the instrument was deployed as a reference base station for several days in Northern Quebec. Reference stations are critical for removing diurnal variations from survey data, ensuring high-fidelity measurements. During testing, the SBQDM operated continuously between –13 °C and 0 °C for up to 20 hours, confirming that the diamond sensing head, optics, and electronics can tolerate severe cold. Although the device produced valid magnetic readings, the bias-field-induced offsets had not been characterized for such low temperatures, resulting in larger day-to-day variations.

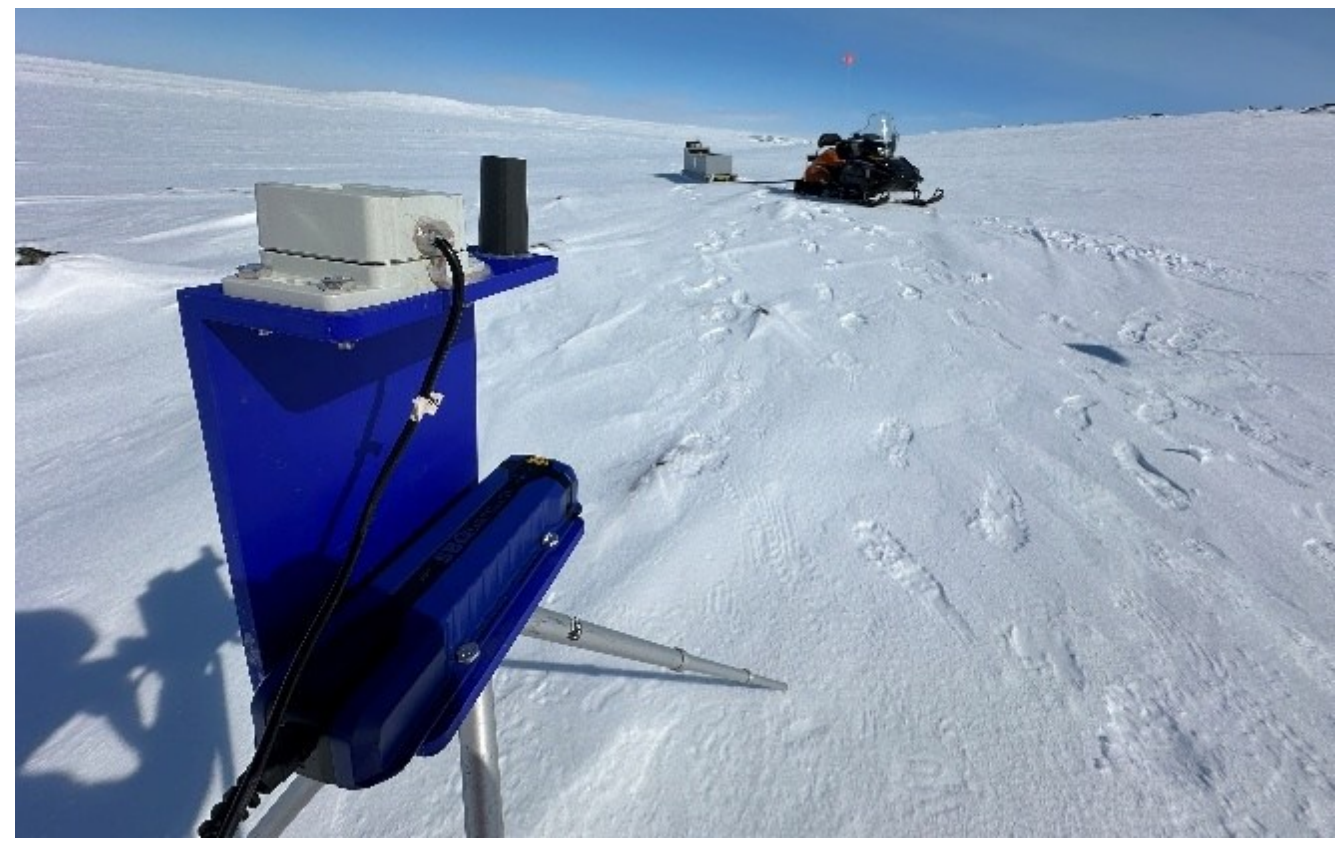

*Figure 10: SBQDM as a static base station in the Canadian North. The snowmobile in the background is pulling other sets of magnetometers for magnetic cartography purposes. The base station signal will allow to remove diurnal variations from the maps.*

Notably, the testing coincided with the major solar storm of May 10th, 2024, which generated field variations on the order of hundreds of nT, clearly captured by the SBQDM (Figure 11). This dataset demonstrates the value of a vector reference base station, as each axis recorded distinct perturbations, highlighting how vector data improves correction of survey results.

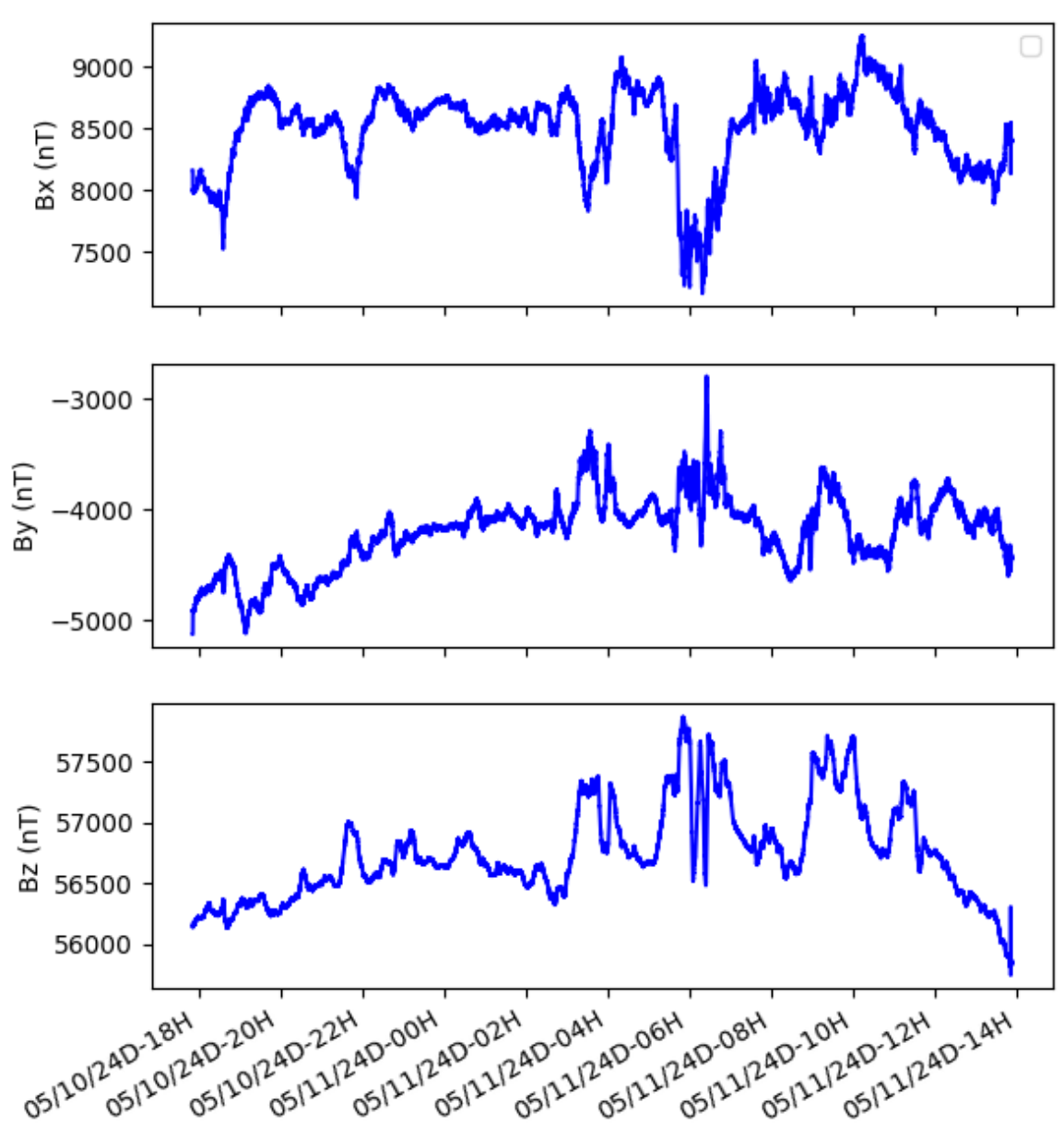

*Figure 11: Magnetic field acquisition in Canadian North, showing variations in the magnetic field due to a solar storm.*

## 4.3 Long-term deployment

At the same time, two SBQDM units were deployed on marble blocks at the NRCan Geophysical Observatory, where they ran continuously for over six months. Their measurements were compared against NRCan's reference magnetometers, providing a long-term assessment of stability, an important feature to validate especially in the context of space deployment where rework and repairs on the device are not an option.

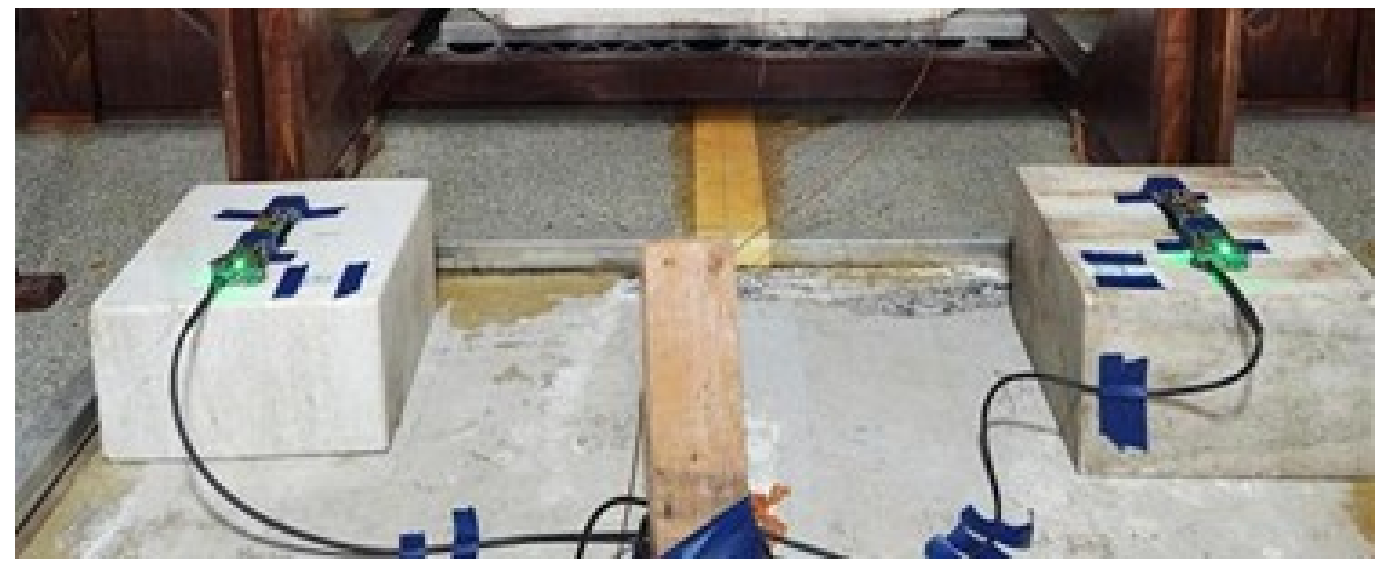

*Figure 12: The two long-term testing SBQDM placed on marble blocks at NRCan Geomagnetic Observatory.*

These calibrated units also captured the May 10th solar storm, showing close agreement with the reference measurements of NRCan. Because calibration was performed two months earlier, the residual offsets observed in the data illustrate the drift of calibration parameters over time.

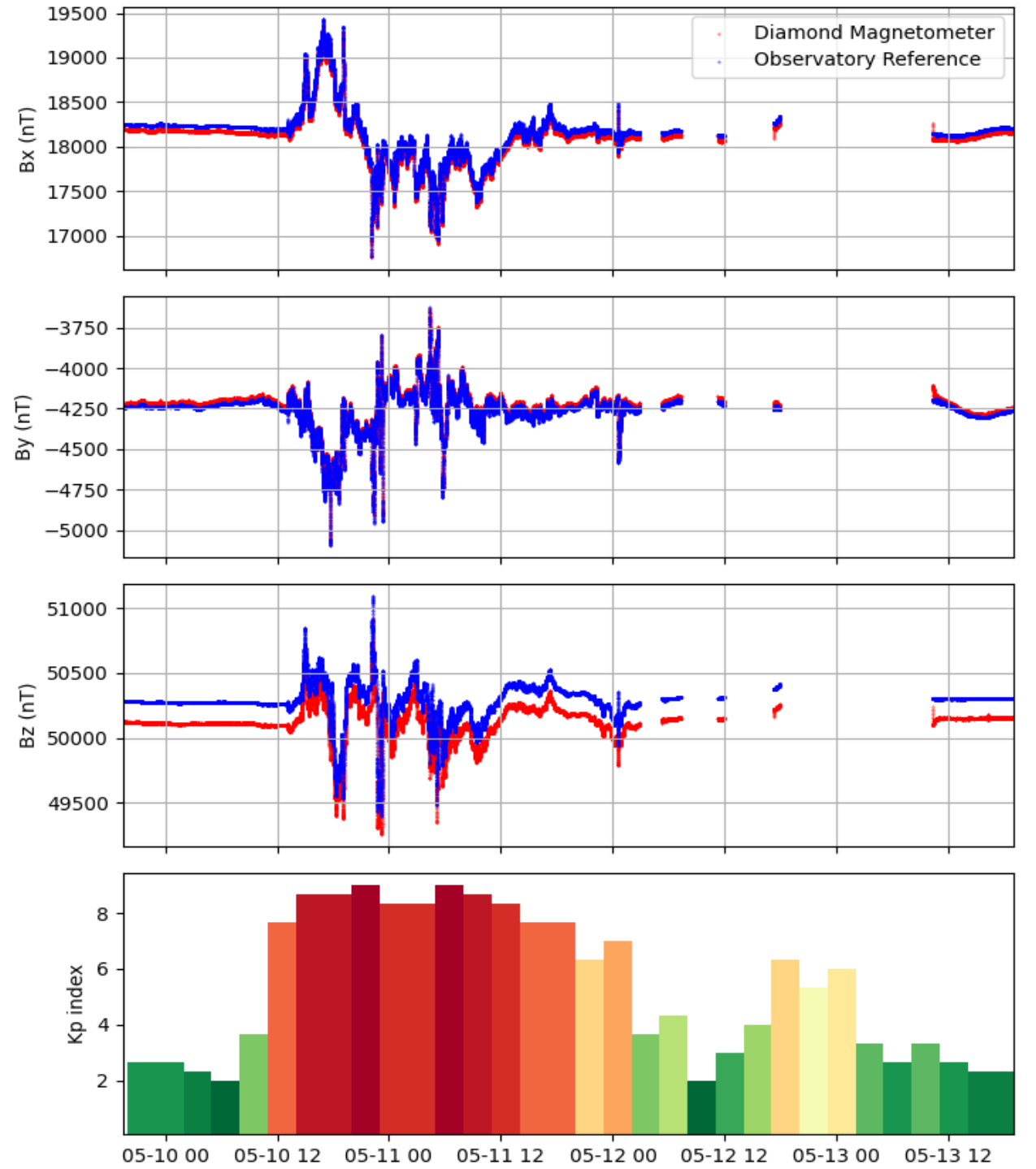

*Figure 13: Magnetic vector signal of the May 10th solar storm of the SBQDM (Reference) in red(blue), with Planetary K-index (Kp index) indicating magnitude of geomagnetic storms 16.*

This extended testing also revealed gradual drift linked to component aging, particularly in a monitoring photodetector used to stabilize laser output. Insights from these long-term campaigns are directly feeding into design iterations to further enhance data quality and long-term stability for future field and space deployments.

## 4.4 Drone deployment

In the mining exploration industry, magnetic surveys are widely used as a rapid method to gain insight into underlying geological formations. Airborne magnetic surveys are particularly valuable, as they can cover large areas quickly and cost-effectively compared to ground-based campaigns [17]. While Total Magnetic Intensity (TMI) maps already provide useful information, they can miss key details, such as remanent magnetization from magnetic minerals, that are better revealed through vector measurements[18]. Having an accurate and sensitive enough vector measurement is a feature that NV center magnetometer can provide, while having low enough SWaP to be embedded onboard small drones. Drones-based magnetometry surveys can fly closer to the ground than traditional airborne surveys, detecting smaller features[19]. Moreover, having a single, accurate vector magnetometer onboard eliminates the need for the traditional combination of scalar and vector sensors used in most airborne vector surveys to compensate for platform magnetic signatures.

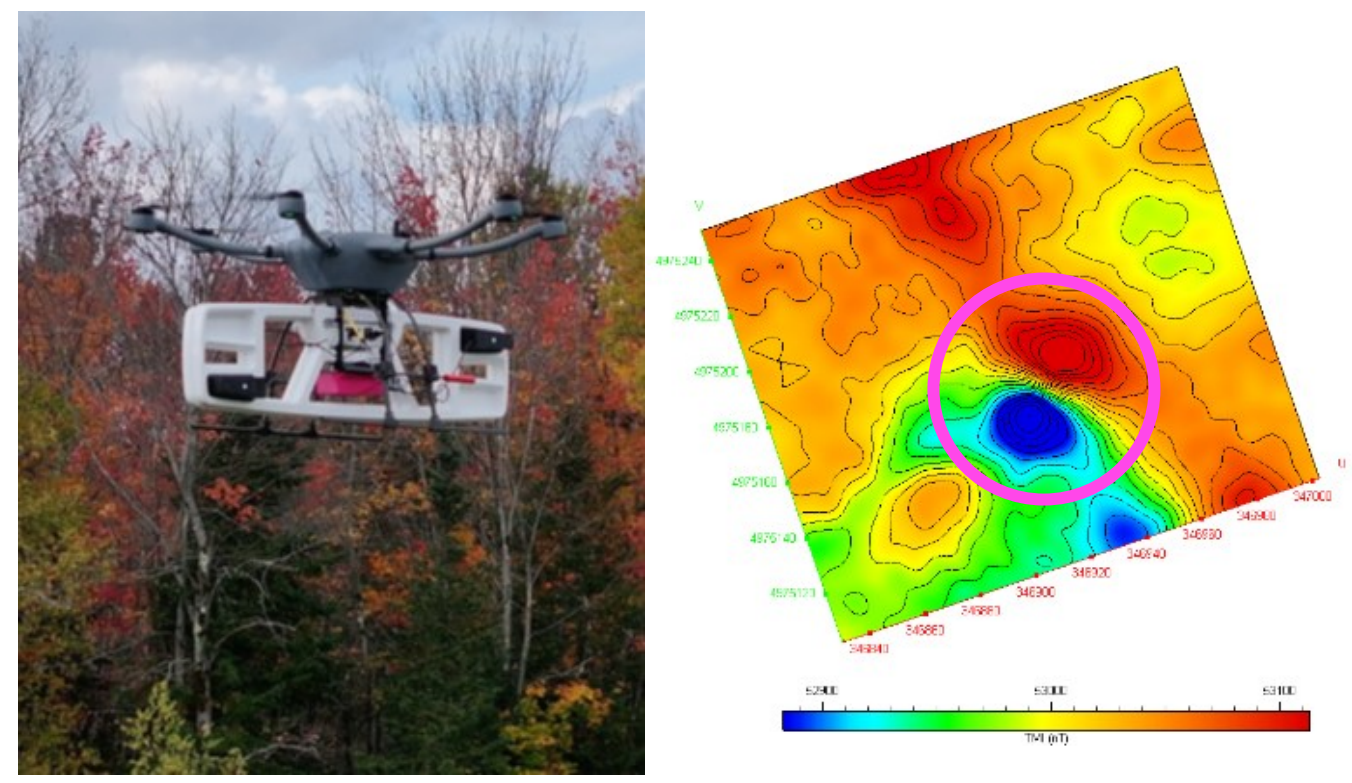

*Figure 14: (left) Drone flight with a SBQDM at a meteorite impact test site. (right) TMI map of the site generated by the combination of the 3 vector components of the SBQDM. The meteorite impact site is highlighted by the pink circle.*

To evaluate the SBQDM's performance in this context, a test survey was conducted at a site in Ontario, Canada, known to contain a magnetic signature from a small meteorite impact. As shown in Figure 14, the SBQDM successfully captured data of sufficient quality to clearly identify the meteorite's magnetic signature, despite the challenges associated with drone integration. The test survey did highlight some sensitivity to mechanical vibration which will be addressed in subsequent designs.

*4.5 Weather balloon deployment*

In June 2024, SBQuantum tested the SBQDM onboard a high-altitude balloon flight launched from the Esrange Space Center, Sweden, as part of the ATMOSFER mission[20]. The purpose of this mission was twofold: to conduct experiments in a near-space environment, characterized by low pressure, extremely low temperatures, limited communication, and exposure to mechanical shocks, while also providing hands-on training for students in quantum technologies.

The SBQDM maintained functionality throughout the whole flight, withstanding temperatures as low as -60 °C at 32 km altitude, and survived the landing impact. These results demonstrate the sensor's robustness and suitability for deployment in extreme environments.

The data acquired highlights the importance of the sensor's bandwidth for vector magnetometry. Due to the balloon's rapid spinning during ascent, confirmed by onboard accelerometer data, the SBQDM recorded highly variable vector readings and noisy Total Magnetic Intensity (TMI) data. Since then, important improvements have been made to the sensor's bandwidth to avoid this problem. We still observe that once the balloon's rotation stabilized near 30 km altitude, the SBQDM produced stable and reliable measurements, confirming its ability to operate effectively at high altitude, even after spinning for two hours. This vector sensitivity to heading is a good example of the use of vector magnetometer for attitude determination.

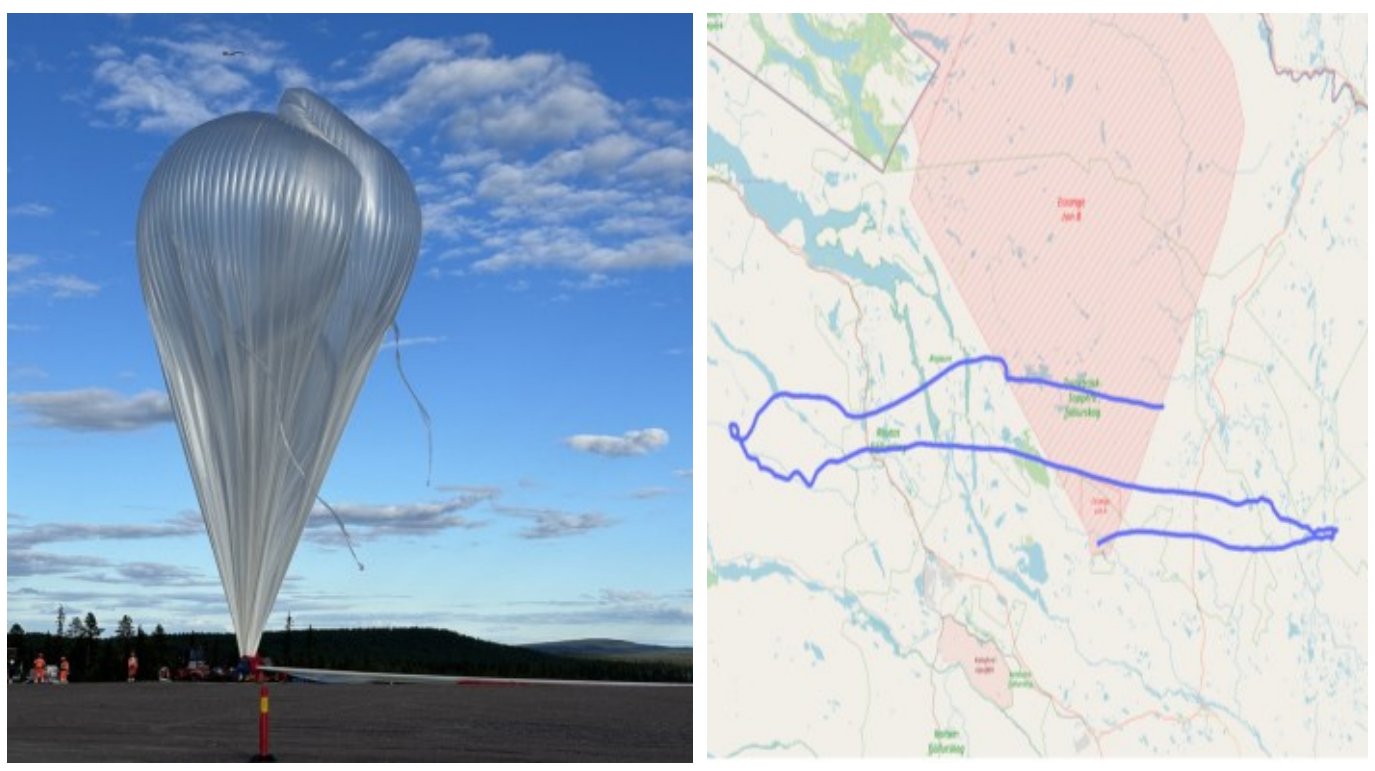

*Figure 15: (left) Stratospheric balloon ready for takeoff with the SBQDM payload. (right) GPS coordinates of the flight path.*

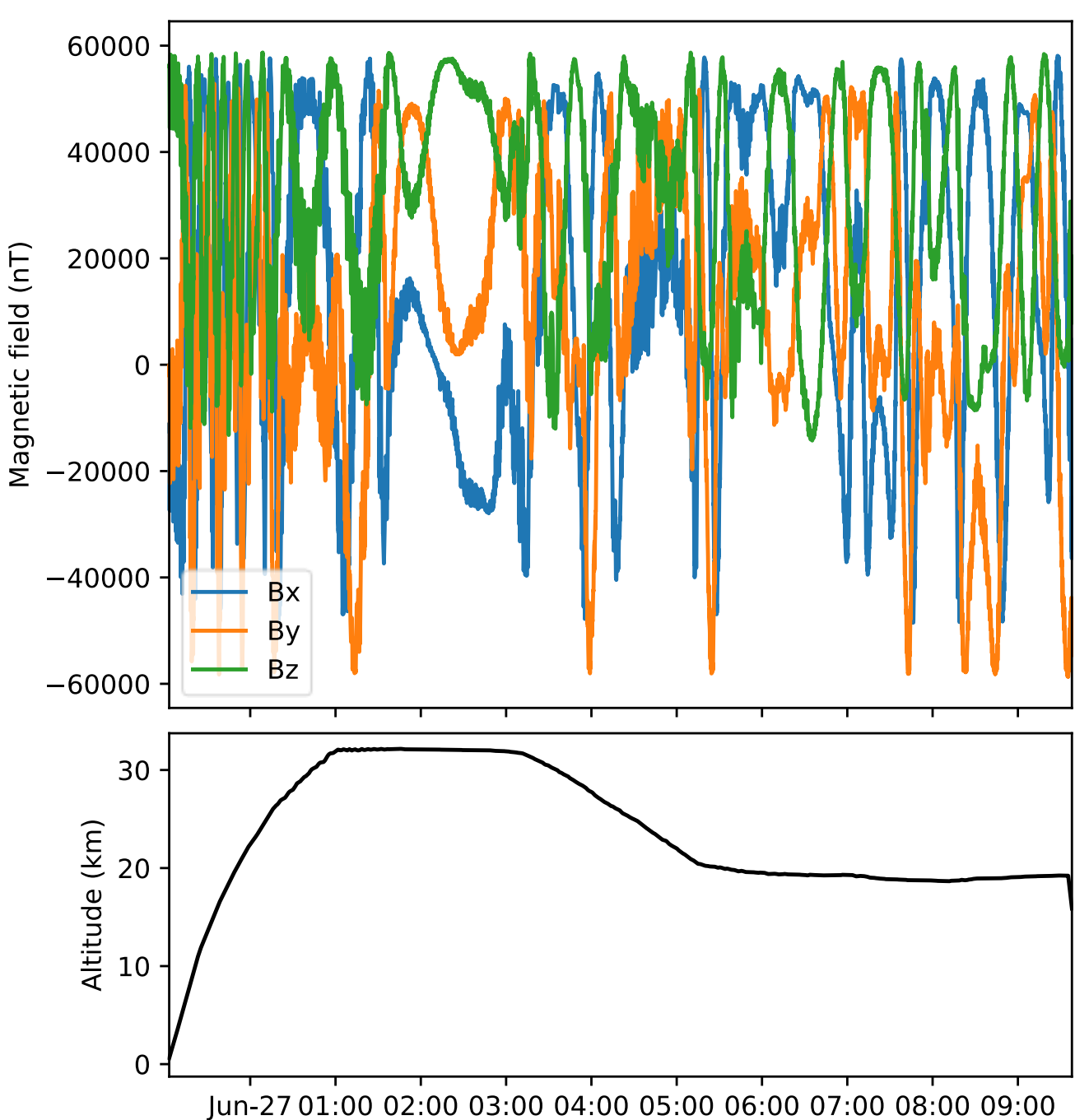

*Figure 16: Magnetic field data acquisition for the ATMOSFER mission, showing periods of instability due to the balloon spinning in ascent and descent.*

## 5. Outlook

The journey of building this magnetometer for space deployment started in 2020 and is about to reach its completion with the launch of the SBQDM payload in early 2026. We demonstrate however in this paper that the usefulness of diamond magnetometer extends far beyond space exploration, and that the potential of absolute vector accuracy of this technology can bring new use cases or improve current ones.

Future developments will target a sensitivity of $50\ pT/\sqrt{Hz}$ and sub-nanotesla stability, achieved by eliminating bias magnets in favor of more stable biasing techniques. The next-generation design will also integrate lessons learned from field deployments, such as improving vibration resilience and increasing bandwidth. Achieving this level of performance means the sensor will operate at or below typical environmental magnetic noise levels, making it truly ready for wider deployment in real-world conditions. Moreover, the enhanced stability will enable the use of arrays of sensors for magnetic intelligence, simplifying magnetic interpretations for end users and improving the classification and mapping of magnetic signals, unlocking SBQuantum's missions to "Reveal the invisible".

**Acknowledgement**

The development of the magnetometer and its tests were funded in part by Canadian Space Agency (22STDPT07, 24STDPV27),

Ministère de l'Économie, de l'Innovation et de l'Énergie (PI70805, PI64436), National Research Council of Canada Quantum Sensors Challenge Program (QSP 051). We want to thank Element Six for the supply of diamond material, Natural Resources Canada for the support and guidance on calibration of magnetometers at their facility, CEGEP de Sherbrooke for collaboration work on the ATMOSFER mission. We want to acknowledge the work of Frédérik Coulombe, Hubert Dubé, Benjamin Dupuis, Isaac Fiset, Léa Lavoie, Andrew Lowther, Julien Poirier-Teasdale, Amin Rezaei and Gabriel St-Hilaire, for contribution to the sensor's performances and demonstrations. Finally, we want to acknowledge the contribution of numerous interns to the sensor's development and field deployment.